# Discrete space-time symmetry, polarization eigenmodes and their degeneracies

R J Potton

Magdalene College, Cambridge

**Abstract**

The irreducible representations of the group $C_4 \otimes V$ can be used to distinguish polarization eigenmodes, to account for their degeneracies and to associate them with particular magnetic crystal classes. The occurrence of this finite Abelian group points to a possible connection with topology but, irrespective of this, qualitative features of gyrotropy in condensed media can be approached in a way that does not depend on arbitrarily truncated tensor expansions.

rjp73@cam.ac.uk

# 1 Introduction

The ability to synthesize to order specimens having optical properties that are present only sporadically, if at all, in naturally occurring materials (1) relies on the exploitation of the fundamental space-time symmetries that ultimately constrain observable phenomena. One phenomenon that still stretches the scope of theoretical analysis is that of gyrotropy. The extent to which circularly polarized modes are or are not degenerate with respect to direction of propagation along any particular spatial axis in a crystal is of practical as well as theoretical interest (2). In particular the usefulness of nonreciprocal devices places a premium on the understanding of the roles of chirality and magnetism in reciprocity. Svirko et al (3) have pointed to the role of PT symmetry at the heart of gyrotropy in crystals. Optical activity is generally attributed to non-local spatial response (4). At the same time the association of Faraday birefringence with magnetism points to its being non-local in time.

Some of these ideas seem already to have been in the minds of Brown et. al. (5) when they classified birefringence phenomena according to their transformation properties under space inversion, P, and time inversion T. However, they did not explicitly invoke any PT symmetry element. Moreover, their argument was couched in terms of property tensors of particular (low) ranks for specific crystal classes, rather than involving any overarching space-time symmetry. Nevertheless, Brown's paper strongly suggests a classification of crystal property tensors responsible for different birefringence effects according to irreducible representations of the Vierergruppe V (6) composed of the set {1,P,T,PT} as shown in figure 1. The group is defined by its being generated by the two elements P and T, space inversion and time inversion.

| V | 1 | T | P | PT | tensor type | phenomenon |
|---|---|---|---|----|-------------|------------|
| $A_g$ | 1 | 1 | 1 | 1 | | |
| $B_g$ | 1 | -1 | 1 | -1 | $2^{nd}$ rank c-tensor | Faraday effect |
| $A_u$ | 1 | 1 | -1 | -1 | $3^{rd}$ rank i-tensor | optically activity |
| $B_u$ | 1 | -1 | -1 | 1 | | |

Figure 1 Character table of the Vierergruppe, V, of which Brown et. al.'s classification of property tensor origins of gyrotropic effects is suggestive.

# 2 Anti-unitary time-reversal

The introduction of time-reversal into quantum theory forces the augmentation of unitary transformations that are sufficient to account for spatial symmetry, with anti-unitary transformations (7). The development of the theory is based on the realisation that unitary symmetry is too restrictive to account for observed degeneracies of states and transition probabilities. Prior to this the symmetries of light polarization phenomena had already been addressed (8). However, anti-unitary symmetries unlike unitary ones do not sit comfortably (9) in the framework of representation space and its dual to which the Hermitian inner product belongs. Adjointness does not have the straightforward meaning for anti-unitary transformations that it does for unitary ones. Moreover, the need to introduce more than one type of co-representation to handle symmetry groups containing anti-unitary operators tends to obscure the essential simple idea that, to quote Wigner, that "It is a consequence of the involutional nature of the physical operation of time inversion that the succession of two time inversions restores the original state" (7a).

The complications ensuing from the introduction of anti-unitary time-reversal are required for matrix elements to be correctly handled in computing transition probabilities. In the next sections it will be shown that the more modest objectives of labelling states, predicting degeneracies of states and formulating selection rules can be achieved without reference to an inner product or the dual space that belongs with it.

## 2 Anti-linear time-reversal

If the object is to identify stationary states and their degeneracies it is possible to avoid dual spaces and co-representations altogether. In the interest of simplicity it is worthwhile to enquire how far it is possible to go without introducing to a representation space an inner product with its associated complexities. The answer seems to be: further than might be supposed. It turns out that, without recourse to an inner product or to any dual space, labelling of states, prediction of degeneracies of states and of selection rules are all accessible. All of these things can be recognised from the linear independence of state functions classified according to irreducible representations of the symmetry group. What is lacking is the possibility of predicting transition probabilities when transitions between states are to be quantified. The gain is that it allows the representations of a symmetry group containing anti-linear time-reversal to be focused on and the origin of consequent degeneracies to be clarified. This is especially useful in analysing the scope of the reciprocity principle in polarization optics (10).

The time–reversal operation, T, is first considered in the anti-linear form:

$$T \sum c_n \xi_n(t) = \sum c_n{}^* T \xi_n(t) \qquad (1)$$

Given the involutory or anti-involutory (9a) character of T we may take the following route to finding irreducible group representations to which states belong. Begin by observing that the power of representation theory in labelling states, predicting their degeneracy and discovering selection rules is not exploited by adopting the obvious matrix representation {1,-1} of the group{1, T} of an involutory T. Specifically, to identify degeneracies we require a symmetry group with more than one-dimensional irreducible representations or, at least, one-dimensional complex conjugate representations. The latter may well be sufficient for the present case as we are dealing with quanta (photons) that although having integer spin are massless and have spin $\pm 1$ only (11). Next observe that the cyclic group of order four has one complex representation. If this is the group that we require – what is its generator? Evidently not T if we are to take Wigner's quotation at face value. On the other hand there are reasons for believing that time and space in the vicinity of matter are separately orientable (12). If this is the case re-orientation of (time or space) followed by reorientation of (space or time) constitutes not an identity but a total reorientation of space-time. On this view re-orientation of (time or space) as a partial reorientation of space-time is an anti-involution – a symmetry operation of order four. This is the premise for what follows with partial re-orientation of space-time designated $I_{ts}$ to emphasise that it is not to be decomposed into more primitive factors. To get a sufficient description of space-time symmetry one must complement these orientation considerations with specific reference to time-inversion and space-inversion. This can be done by taking as the symmetry group the direct product of $C_4$ with V (the generators of which will be designated $I_t$ and $I_s$).

For the purposes of this paper the time reversal operation is taken to be:

$$I_t : \sum c_n \psi_n(t) \mapsto \sum c_n * \psi_n(-t) \qquad (2)$$

while space inversion is the involution:

$$I_s : \chi(\mathbf{r}) \mapsto \chi(-\mathbf{r}) \qquad (3)$$

## 3 Linear time inversion

Complex conjugation in * algebras may be viewed in different ways that are usually considered to be equivalent . In this section it will be argued that complex analysis without conjugation has some utility in relation to the analysis of discrete space-time symmetries. To a mathematician the essence of complex conjugation is perhaps the automorphism $i \leftrightarrow -i$ of the multiplicative group {1, i, -1, -i} or a change of the sense of the contour in Cauchy's integral formula. Otherwise, reflections in the real axis of the complex plane are $b \mapsto -b$ in the Cartesian, a + bi, or $\phi \mapsto -\phi$ in the polar, $|z|\exp(i\phi)$, representation of a complex number z. Why might one wish to distinguish these interpretations of z* ? To the engineer, complex conjugation brings to mind the phasor representation of harmonic disturbances, while to the optical physicist it suggests complex representation of progressive waves with phase incorporated in the complex amplitude. In this context the polar angle, $\phi$, may be interpreted as the phase $\mathbf{k.r} - \omega t$ of a monochromatic wave. In this case a connection with space-time symmetries is clearly indicated. It will be argued that, in these cases, the dual mapping $z \mapsto z*$ is far from itself being a symmetry (vide analytic continuation in the upper half complex plane). Nevertheless, apart from any utility in labelling circularly (or indeed elliptically) polarized states and accounting for their degeneracies there must be some more fundamental justification for the effort required to deconstruct symmetry groups with anti-unitary elements and force time-reversal into a linear framework. This is believed to lie in the fact that the conjugate mapping is misleading in obscuring the basis for time-reversal symmetry in orientation properties in space-time the proper vehicle for which is homology theory.

A transverse wave $\begin{pmatrix} c_1 \\ c_2 \end{pmatrix} \exp[i(kz - \omega t)]$ has transverse field components $E_x$ and $E_y$ represented by complex amplitudes $c_1 = |c_1|\exp(i\theta_1)$ and $c_2 = |c_2|\exp(i\theta_2)$. In the case that $\arg\left(\dfrac{c_1}{c_2}\right) = \theta_1 - \theta_2 = \mp\dfrac{\pi}{2}$ it follows that $\arg\left(\dfrac{c_1*}{c_2*}\right) = \theta_2 - \theta_1 = \pm\dfrac{\pi}{2}$ can be written as $\arg\left(\dfrac{\pm ic_1}{\mp ic_2}\right) = \arg\left(\dfrac{c_1*}{c_2*}\right)$. This correspondence opens the way to the use of $2 \times 2$ matrix, linear transformations, $\begin{bmatrix} \pm i & 0 \\ 0 & \mp i \end{bmatrix}$ to handle the otherwise dual operation of time-reversal of disturbances in quadrature. Furthermore, if $\begin{bmatrix} i & 0 \\ 0 & -i \end{bmatrix}$ yields the conjugate of the vector $\begin{pmatrix} c_1 \\ c_2 \end{pmatrix}$ its inverse yields the ray equivalent of the conjugate. These linear operators allow the degenerate functions that form the bases of irreducible representations of a finite symmetry group to be identified. Thus, complex conjugation represented by a cyclic group of order four generated by $\begin{bmatrix} i & 0 \\ 0 & -i \end{bmatrix}$ amounts, as a ray representation, to an involution. When combined with the other part of the time-reversal operation, namely the mapping $t \mapsto -t$, and with space inversion $\mathbf{r} \mapsto -\mathbf{r}$ the means is established for classifying states according to their discrete space-time symmetries.

The progressive wave character of $\begin{pmatrix} c_1 \\ c_2 \end{pmatrix} \exp[i(\mathbf{k}.\mathbf{r} - \omega t)]$ is contained in $\exp[i(\mathbf{k}.\mathbf{r} - \omega t)] = \exp(i\phi)$ and this will now be put into a form in which there is manifest symmetry, in the form of linear transformations, under $t \mapsto -t$ and $\mathbf{r} \mapsto -\mathbf{r}$. The imaginary part of $\exp(i\phi)$ can be projected as:

$$\sin(\phi) = \frac{\exp(i\phi) - \exp(-i\phi)}{2i} \tag{4}$$

using $i \mapsto -i$, or just as well, by $\phi \mapsto -\phi$. Thus, the conjugate equivalent imposed on amplitudes $\begin{pmatrix} c_1 \\ c_2 \end{pmatrix}$ yields the transformation $\sin(\phi) \mapsto -\sin(\phi)$ of such a wavefunction. On the other hand the separate symmetries $t \mapsto -t$ and $\mathbf{r} \mapsto -\mathbf{r}$ yield linearly independent functions in acting upon $\sin(\phi) = \sin(\mathbf{k}.\mathbf{r} - \omega t)$. Provided that the group of the symmetries $\phi \mapsto -\phi$, $t \mapsto -t$ and $\mathbf{r} \mapsto -\mathbf{r}$ can be identified, we have a means of identifying the degeneracies of circularly (or elliptically) polarized waves.

The group in question is $C_4 \otimes V$. $C_4$ is the cyclic group of order four generated by $I_{ts} : \phi \mapsto -\phi$ already identified as partial re-orientation of space-time. V is the Vierergruppe mentioned above, generated by separate time inversion, $I_t : t \mapsto -t$, and space inversion, $I_s : \mathbf{r} \mapsto -\mathbf{r}$. The character table for $C_4 \otimes V$ is shown in figure 2. Given that it is commutative the multiplication of group elements is apparent from the labelling of columns in its character table.

That the group $C_4 \otimes V$ is Abelian is significant. The behaviour of gyrotropic electromagnetic modes in matter ultimately has to do with orientation in space-time the mathematics of which is homology and in homology Abelian groups have a key role.

| $C_4 \otimes V$ | 1 | $I_{ts}$ | $I_{ts}^2$ | $I_{ts}^3$ | $I_t$ | $I_tI_{ts}$ | $I_tI_{ts}^2$ | $I_tI_{ts}^3$ | $I_s$ | $I_sI_{ts}$ | $I_sI_{ts}^2$ | $I_sI_{ts}^3$ | $I_tI_s$ | $I_tI_sI_{ts}$ | $I_tI_sI_{ts}^2$ | $I_tI_sI_{ts}^3$ |
|---|---|---|---|---|---|---|---|---|---|---|---|---|---|---|---|---|
| $A_{gi}$ | 1 | 1 | 1 | 1 | 1 | 1 | 1 | 1 | 1 | 1 | 1 | 1 | 1 | 1 | 1 | 1 |
| $B_{gi}$ | 1 | -1 | 1 | -1 | 1 | -1 | 1 | -1 | 1 | -1 | 1 | -1 | 1 | -1 | 1 | -1 |
| $E_{gi}$ | 1 | i | -1 | -i | 1 | i | -1 | -i | 1 | i | -1 | -i | 1 | i | -1 | -i |
|  | 1 | -i | -1 | i | 1 | -i | -1 | i | 1 | -i | -1 | i | 1 | -i | -1 | i |
| $A_{ui}$ | 1 | 1 | 1 | 1 | 1 | 1 | 1 | 1 | -1 | -1 | -1 | -1 | -1 | -1 | -1 | -1 |
| $B_{ui}$ | 1 | -1 | 1 | -1 | 1 | -1 | 1 | -1 | -1 | 1 | -1 | 1 | -1 | 1 | -1 | 1 |
| $E_{ui}$ | 1 | i | -1 | -i | 1 | i | -1 | -i | -1 | -i | 1 | i | -1 | -i | 1 | i |
|  | 1 | -i | -1 | i | 1 | -i | -1 | i | -1 | i | 1 | -i | -1 | i | 1 | -i |
| $A_{gc}$ | 1 | 1 | 1 | 1 | -1 | -1 | -1 | -1 | 1 | 1 | 1 | 1 | -1 | -1 | -1 | -1 |
| $B_{gc}$ | 1 | -1 | 1 | -1 | -1 | 1 | -1 | 1 | 1 | -1 | 1 | -1 | -1 | 1 | -1 | 1 |
| $E_{gc}$ | 1 | i | -1 | -i | -1 | -i | 1 | i | 1 | i | -1 | -i | -1 | -i | 1 | i |
|  | 1 | -i | -1 | i | -1 | i | 1 | -i | 1 | -i | -1 | i | -1 | i | 1 | -i |
| $A_{uc}$ | 1 | 1 | 1 | 1 | -1 | -1 | -1 | -1 | -1 | -1 | -1 | -1 | 1 | 1 | 1 | 1 |
| $B_{uc}$ | 1 | -1 | 1 | -1 | -1 | 1 | -1 | 1 | -1 | 1 | -1 | 1 | 1 | -1 | 1 | -1 |
| $E_{uc}$ | 1 | i | -1 | -i | -1 | -i | 1 | i | -1 | -i | 1 | i | 1 | i | -1 | -i |
|  | 1 | -i | -1 | i | -1 | i | 1 | -i | -1 | i | 1 | -i | 1 | -i | -1 | i |

Figure 2 Character table for the Abelian group $C_4 \otimes V$. The representation labels have subscripts g and u, standing for gerade and ungerade, and i and c, standing for (time) invariant and (time) change.

## 4 Mode transformations and degeneracies

Functions like $\begin{bmatrix} 1+i \\ 1-i \end{bmatrix} \sin(kz - \omega t)$ provide bases for representations of $C_4 \otimes V$ in the form of Cartesian products ($\times$) of two-dimensional complex vectors and real functions of **r** and t. Figure 3 shows the results of the various group actions of $C_4 \otimes V$ acting on a mode $\begin{bmatrix} 1+i \\ 1-i \end{bmatrix} \sin(kz - \omega t)$ in the $E_{ui} \times B_{ui}$ basis. It can be seen that, as expected there is degeneracy between right forward and right backward modes.

Left forward and left backward modes together form another, linearly independent, basis ($E_{ui} \times B_{ui}$)'. Unless by reason of some specific material symmetry the left and right modes are non-degenerate.

However, some explanation is required of how these results are arrived at. The crucial question of whether or not $E_y$ leads $E_x$ requires a consideration of the complementary actions of * (or its surrogate $\begin{bmatrix} i & 0 \\ 0 & -i \end{bmatrix}$) and $t \mapsto -t$. This is set out in figure 4.

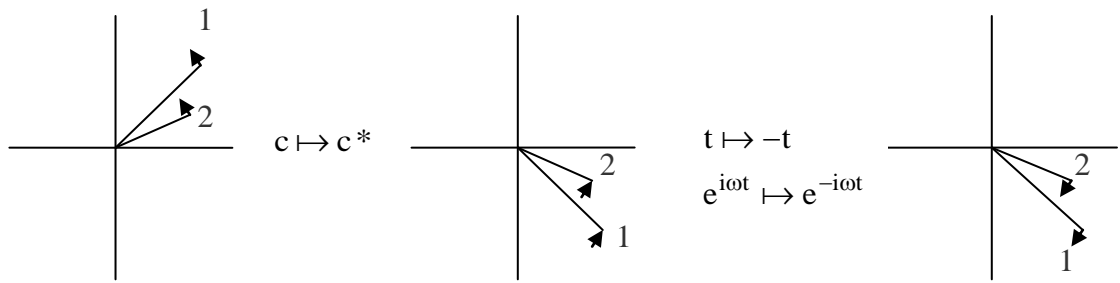

Figure 4 A leading/lagging relationship between 1 and 2 is preserved by anti-linear time reversal.

According to figure 4, either $c \mapsto c*$ or $t \mapsto -t$ on their own convert a leading to a lagging phase relationship. Together they leave the leading lagging relationship unchanged and constitute a rather generally applicable type of symmetry.

| operation | state function | φ' | leading | direction | mode |
|---|---|---|---|---|---|
| $I = \begin{bmatrix} 1 & 0 \\ 0 & 1 \end{bmatrix}$ | $\begin{bmatrix} 1+i \\ 1-i \end{bmatrix} \sin(kz - \omega t)$ | $\phi$ | $E_y$ | positive z | right +z |
| $I_{ts} = \begin{bmatrix} i & 0 \\ 0 & -i \end{bmatrix}$ | $\begin{bmatrix} -1+i \\ -1-i \end{bmatrix} \sin(-kz + \omega t)$ | $-\phi$ | $E_y$ | positive z | right +z |
| $I_{ts}^2 = \begin{bmatrix} -1 & 0 \\ 0 & -1 \end{bmatrix}$ | $\begin{bmatrix} -1-i \\ -1+i \end{bmatrix} \sin(kz - \omega t)$ | $\phi$ | $E_y$ | positive z | right +z |
| $I_{ts}^3 = \begin{bmatrix} -i & 0 \\ 0 & i \end{bmatrix}$ | $\begin{bmatrix} 1-i \\ 1+i \end{bmatrix} \sin(-kz + \omega t)$ | $-\phi$ | $E_y$ | positive z | right +z |
| $I_t = \begin{bmatrix} 1 & 0 \\ 0 & 1 \end{bmatrix}$ | $\begin{bmatrix} 1+i \\ 1-i \end{bmatrix} \sin(kz + \omega t)$ | $\phi$ | $E_x$ | negative z | right -z |
| $I_t I_{ts} = \begin{bmatrix} i & 0 \\ 0 & -i \end{bmatrix}$ | $\begin{bmatrix} -1+i \\ -1-i \end{bmatrix} \sin(-kz - \omega t)$ | $-\phi$ | $E_x$ | negative z | right -z |
| $I_t I_{ts}^2 = \begin{bmatrix} -1 & 0 \\ 0 & -1 \end{bmatrix}$ | $\begin{bmatrix} -1-i \\ -1+i \end{bmatrix} \sin(kz + \omega t)$ | $\phi$ | $E_x$ | negative z | right -z |
| $I_t I_{ts}^3 = \begin{bmatrix} -i & 0 \\ 0 & i \end{bmatrix}$ | $\begin{bmatrix} 1-i \\ 1+i \end{bmatrix} \sin(-kz - \omega t)$ | $-\phi$ | $E_x$ | negative z | right -z |
| $I_s = \begin{bmatrix} -1 & 0 \\ 0 & -1 \end{bmatrix}$ | $\begin{bmatrix} -1-i \\ -1+i \end{bmatrix} \sin(kz + \omega t)$ | $-\phi$ | $E_x$ | negative z | right -z |
| $I_s I_{ts} = \begin{bmatrix} -i & 0 \\ 0 & i \end{bmatrix}$ | $\begin{bmatrix} 1-i \\ 1+i \end{bmatrix} \sin(-kz - \omega t)$ | $\phi$ | $E_x$ | negative z | right -z |
| $I_s I_{ts}^2 = \begin{bmatrix} 1 & 0 \\ 0 & 1 \end{bmatrix}$ | $\begin{bmatrix} 1+i \\ 1-i \end{bmatrix} \sin(kz + \omega t)$ | $-\phi$ | $E_x$ | negative z | right -z |
| $I_s I_{ts}^3 = \begin{bmatrix} i & 0 \\ 0 & -i \end{bmatrix}$ | $\begin{bmatrix} -1+i \\ -1-i \end{bmatrix} \sin(-kz - \omega t)$ | $\phi$ | $E_x$ | negative z | right -z |
| $I_t I_s = \begin{bmatrix} -1 & 0 \\ 0 & -1 \end{bmatrix}$ | $\begin{bmatrix} -1-i \\ -1+i \end{bmatrix} \sin(kz - \omega t)$ | $-\phi$ | $E_y$ | positive z | right +z |
| $I_t I_s I_{ts} = \begin{bmatrix} -i & 0 \\ 0 & i \end{bmatrix}$ | $\begin{bmatrix} 1-i \\ 1+i \end{bmatrix} \sin(-kz + \omega t)$ | $\phi$ | $E_y$ | positive z | right +z |
| $I_t I_s I_{ts}^2 = \begin{bmatrix} 1 & 0 \\ 0 & 1 \end{bmatrix}$ | $\begin{bmatrix} 1+i \\ 1-i \end{bmatrix} \sin(kz - \omega t)$ | $-\phi$ | $E_y$ | positive z | right +z |
| $I_t I_s I_{ts}^3 = \begin{bmatrix} i & 0 \\ 0 & -i \end{bmatrix}$ | $\begin{bmatrix} -1+i \\ -1-i \end{bmatrix} \sin(-kz + \omega t)$ | $\phi$ | $E_y$ | positive z | right +z |

Figure 3 group actions in the $E_{ui} \times B_{ui}$ basis of $C_4 \otimes V$

Figure 6 in the appendix shows the corresponding group actions in the $E_{gc} \times B_{gc}$ basis of $C_4 \otimes V$.

A summary of the classification and degeneracies of mode according to irreducible representations of $C_4 \otimes V$ is shown in figure 5. Arago modes are defined as having reciprocity for the same handedness, Faraday modes reciprocity for opposite handedness.

| $C_4 \otimes V$ representation | degeneracies and modes |
|---|---|
| $E_{gi}$ | right forward ↔ left forward<br>not gyrotropic |
| $E_{gc}$ | right forward ↔ left backward<br>Faraday type |
| $E_{ui}$ | right forward ↔ right backward<br>Arago type |
| $E_{uc}$ | Faraday and Arago<br>modes coexist |

Figure 5 mode types and degeneracies by group representation

Of the ninety magnetic point groups (13) those in classes identified by Birss (14) as being of types $n_1$, $m_2$ and $m_3$ transform respectively as representations $E_{gi}$, $E_{gc}$ and $E_{uc}$. Each type exhibits characteristic electromagnetic mode polarization and degeneracy properties. Together these account for fifty three of the ninety classes. Of the remaining thirty seven, twenty one transform like $E_{ui}$ (the nonmagnetic noncentrosymmetric crystal classes) while sixteen belong to magnetic classes that have neither $I_t$ nor $I_s$ nor yet $I_t I_s$ symmetry. All of the thirty seven are circularly birefringent but only those belonging to $E_{ui}$ exhibit degeneracy.

For symmetric classes belonging to $E_{uc}$ the argument adduced here breaks down. It predicts degeneracy of right and left forward modes. On the face of it this might be thought to indicate absence of gyrotropy. However, the representation is linearly independent of $E_{gi}$ (to which the nonmagnetic centrosymmetric classes belong) and another possibility exists. This is that there is a nondegenerate linearly independent though similarly polarized mode in the forward direction belonging to a second basis for $E_{uc}$. That there may exist degeneracy between forward and backward modes in this case poses a question about the adequacy of the phasor diagram argument to account for what are ultimately orientated entities in space-time.

# 5 Reciprocity and non-reciprocity from circular birefringence

The fact that gyrotropic modes may belong to different symmetry group irreducible representations suggests the possibility that Arago modes and Faraday modes may coexist in the same medium. The circularly polarised modes realised in ellipsometry will be linear combinations of the two types. These modes having nearly equal wave impedance mismatch at the entrance surface will be excited with nearly equal amplitude. Even for normally incident circularly polarised light in the direction of a symmetry axis of the crystal and in the absence of linear birefringence, an eigenmode expansion in terms of an Arago mode and a Faraday mode will be required. The Arago mode will beat with the Faraday mode of the same handedness and the result after a long enough propagation path will be a major change of polarisation state. On reflection, the Arago mode will beat with the Faraday mode of opposite handedness, the beat length being different. In this sense reciprocity of propagation is broken as has been observed by (10). To excite separately each type of circularly polarised mode it will be necessary to employ velocity coupling (using grating coupling or attenuated total reflection prism coupling) as opposed to end-fire launching so as to discriminate between eigenmodes by way of their phase speeds.

# 6 Conclusion

An attempt has been made to account for the degeneracies of polarization eigenmodes in terms of the irreducible representations of a finite group of discrete space-time symmetries, $C_4 \otimes V$. Of the four complex irreducible representations of this group the one associated with centrosymmetric nonmagnetic crystals has as basis functions plane polarized electromagnetic modes. The other three, each associated with a particular set of magnetic symmetry classes, label media that exhibit i) optical activity ii) Faraday effect and iii) both optical activity and Faraday effect. Crystals that are unsymmetrical are gyrotropic but exhibit no nonaccidental degeneracy. Hence, the paper does provide a description of circular birefringence that is free from the myriad particularities of coloured point and space groups. However, it barely hints at a more seamless account of gyrotropy that must surely include topological invariants.

**Appendix**

| operation | state function | $\phi'$ | leading | direction | mode |
|---|---|---|---|---|---|
| $I = \begin{bmatrix} 1 & 0 \\ 0 & 1 \end{bmatrix}$ | $\begin{bmatrix} 1+i \\ 1-i \end{bmatrix} \sin(kz - \omega t)$ | $\phi$ | $E_y$ | positive z | right +z |
| $I_{ts} = \begin{bmatrix} i & 0 \\ 0 & -i \end{bmatrix}$ | $\begin{bmatrix} -1+i \\ -1-i \end{bmatrix} \sin(-kz + \omega t)$ | $-\phi$ | $E_y$ | positive z | right +z |
| $I_{ts}^2 = \begin{bmatrix} -1 & 0 \\ 0 & -1 \end{bmatrix}$ | $\begin{bmatrix} -1-i \\ -1+i \end{bmatrix} \sin(kz - \omega t)$ | $\phi$ | $E_y$ | positive z | right +z |
| $I_{ts}^3 = \begin{bmatrix} -i & 0 \\ 0 & i \end{bmatrix}$ | $\begin{bmatrix} 1-i \\ 1+i \end{bmatrix} \sin(-kz + \omega t)$ | $-\phi$ | $E_y$ | positive z | right +z |
| $I_t = \begin{bmatrix} -1 & 0 \\ 0 & -1 \end{bmatrix}$ | $\begin{bmatrix} -1-i \\ -1+i \end{bmatrix} \sin(-kz - \omega t)$ | $-\phi$ | $E_y$ | negative z | left -z |
| $I_t I_{ts} = \begin{bmatrix} -i & 0 \\ 0 & i \end{bmatrix}$ | $\begin{bmatrix} 1-i \\ 1+i \end{bmatrix} \sin(kz + \omega t)$ | $\phi$ | $E_y$ | negative z | left -z |
| $I_t I_{ts}^2 = \begin{bmatrix} 1 & 0 \\ 0 & 1 \end{bmatrix}$ | $\begin{bmatrix} 1+i \\ 1-i \end{bmatrix} \sin(-kz - \omega t)$ | $-\phi$ | $E_y$ | negative z | left -z |
| $I_t I_{ts}^3 = \begin{bmatrix} i & 0 \\ 0 & -i \end{bmatrix}$ | $\begin{bmatrix} -1+i \\ -1-i \end{bmatrix} \sin(kz + \omega t)$ | $\phi$ | $E_y$ | negative z | left -z |
| $I_s = \begin{bmatrix} 1 & 0 \\ 0 & 1 \end{bmatrix}$ | $\begin{bmatrix} 1+i \\ 1-i \end{bmatrix} \sin(-kz - \omega t)$ | $\phi$ | $E_y$ | negative z | left -z |
| $I_s I_{ts} = \begin{bmatrix} i & 0 \\ 0 & -i \end{bmatrix}$ | $\begin{bmatrix} -1+i \\ -1-i \end{bmatrix} \sin(kz + \omega t)$ | $-\phi$ | $E_y$ | negative z | left -z |
| $I_s I_{ts}^2 = \begin{bmatrix} -1 & 0 \\ 0 & -1 \end{bmatrix}$ | $\begin{bmatrix} -1-i \\ -1+i \end{bmatrix} \sin(-kz - \omega t)$ | $\phi$ | $E_y$ | negative z | left -z |
| $I_s I_{ts}^3 = \begin{bmatrix} -i & 0 \\ 0 & i \end{bmatrix}$ | $\begin{bmatrix} 1-i \\ 1+i \end{bmatrix} \sin(kz + \omega t)$ | $-\phi$ | $E_y$ | negative z | left -z |
| $I_t I_s = \begin{bmatrix} -1 & 0 \\ 0 & -1 \end{bmatrix}$ | $\begin{bmatrix} -1-i \\ -1+i \end{bmatrix} \sin(kz - \omega t)$ | $-\phi$ | $E_y$ | positive z | right +z |
| $I_t I_s I_{ts} = \begin{bmatrix} -i & 0 \\ 0 & i \end{bmatrix}$ | $\begin{bmatrix} 1-i \\ 1+i \end{bmatrix} \sin(-kz + \omega t)$ | $\phi$ | $E_y$ | positive z | right +z |
| $I_t I_s I_{ts}^2 = \begin{bmatrix} 1 & 0 \\ 0 & 1 \end{bmatrix}$ | $\begin{bmatrix} 1+i \\ 1-i \end{bmatrix} \sin(kz - \omega t)$ | $-\phi$ | $E_y$ | positive z | right +z |
| $I_t I_s I_{ts}^3 = \begin{bmatrix} i & 0 \\ 0 & -i \end{bmatrix}$ | $\begin{bmatrix} -1+i \\ -1-i \end{bmatrix} \sin(-kz + \omega t)$ | $\phi$ | $E_y$ | positive z | right +z |

Figure 6 group actions in the $E_{gc} \times B_{gc}$ basis of $C_4 \otimes V$